\documentclass[journal]{IEEEtran}
\ifCLASSINFOpdf
   \usepackage[pdftex]{graphicx}
  \else
  \usepackage[dvips]{graphicx}

\fi

\usepackage{amsmath}
\usepackage{algorithmic}
\usepackage{algorithm}

\usepackage{array}
\usepackage{subfigure,color}
\usepackage[2]{editing}

\begin{document}
%
\title{Study of Robust Adaptive Beamforming with Covariance Matrix Reconstruction Based on Power Spectral Estimation and Uncertainty Region}
%
%
%

\author{Saeed~Mohammadzadeh$^{\dag}$, V\'{\i}tor~H.~Nascimento$^{\dag}$, Rodrigo~C.~de~Lamare$^{\dag\dag}$ and~Osman~Kukrer$^{\dag\dag\dag}$
}

\maketitle

\begin{abstract}
In this work, a simple and effective robust adaptive beamforming technique is proposed for uniform linear arrays, which is based on the power spectral estimation and uncertainty region (PSEUR) of the interference plus noise (IPN) components. In particular, two algorithms are presented to find the angular sector of interference in every snapshot based on the adopted spatial uncertainty region of the interference direction. Moreover, a power spectrum is introduced based on the estimation of the power of interference and noise components, which allows the development of a robust approach to IPN covariance matrix reconstruction. The proposed method has two main advantages. First, an angular region that contains the interference direction is updated based on the statistics of the array data. Secondly, the proposed IPN-PSEUR method avoids estimating the power spectrum of the whole range of possible directions of the interference sector. Simulation results show that the performance of the proposed IPN-PSEUR beamformer is almost always close to the optimal value across a wide range of signal-to-noise ratios.
\end{abstract}

\begin{IEEEkeywords}
Covariance matrix reconstruction, DoA estimation, Uncertainty of interference region, Robust adaptive beamforming,  Power spectrum estimation.
\end{IEEEkeywords}
%
\IEEEpeerreviewmaketitle
\section{Introduction}
\IEEEPARstart{A}{daptive} beamforming is a spatial filtering technique that has numerous applications in radar, wireless communications, speech signal processing and other fields \cite{godara1997application,van2004detection}. The standard minimum variance distortionless response (MVDR) \cite{capon1969high} beamformer is a well-known adaptive beamformer that assumes accurate knowledge of the antenna array, the actual array manifold and that there is no desired signal component in the training samples. MVDR is used to suppress interference and noise, while keeping a distortionless desired signal. However, in non-ideal conditions such as mismatches between assumed and true steering vectors (SV) and finite data samples, the effectiveness of adaptive beamformers degrades significantly. During the last decades there have been several contributions to enhance beamformer robustness, for example the so-called diagonal loading technique \cite{li2003robust,elnashar2006further, kukrer2014generalised,rstap,rcfprec}, uncertainty set-based methods \cite{vorobyov2003robust,yu2010robust}, eigenspace-based methods \cite{huang2012modified,gbd,wlbd,rsbd}, the modified robust Capon beamformer in \cite{mohammadzadeh2018modified}, and very recently, the adaptive beamformig based on complex-valued convolutional neural networks for sensor arrays in \cite{mohammadzadeh2022letter}. Nonetheless, the presence of a signal of interest (SOI) component in the training data, especially at high SNR, is a main reason of adaptive beamforming performance degradation.

\subsection{Prior and Related Work}

To address the performance degradation related to the presence of a SOI component in adaptive beamforming, various techniques based on interference-plus-noise covariance (IPN) covariance matrix reconstruction have been recently developed to eliminate the SOI component from the signal covariance matrix. In \cite{gu2012robust,chen2015robust}, the IPN covariance matrix is reconstructed using a Capon power spectrum estimator by integrating over a region separated from the SOI. In \cite{gu2014robust}, by exploiting the sparsity of source distribution, the IPN covariance matrix is reconstructed via solving a compressive sensing problem. It performs perfectly for ideal antenna arrays, but it can not cope with array calibration errors. In \cite{ruan2014robust}, \cite{ruan2016} and \cite{lrcc}, computationally efficient algorithms were obtained via low complexity reconstruction of the IPN covariance matrix. A partial power spectrum sampling approach was developed in \cite{zhang2016interference} to reconstruct the IPN covariance matrix with low computer complexity using the covariance matrix taper technique. In \cite{yuan2017robust}, the IPN covariance matrix is reconstructed by searching for interference steering vectors that lie inside the intersection of the signal-interference subspace and the interference subspace.

The IPN covariance matrix reconstruction in \cite{zheng2018covariance} follows a similar technique to that in \cite{gu2012robust} and \cite{li2003robust}. However, the interference steering vector is estimated based on ad-hoc parameters. The approach in \cite{mohammadzadeh2018adaptive} employs the beamformer output power to jointly estimate the theoretical IPN covariance matrix and the mismatched steering vector using the eigenvalue decomposition of the received signal covariance matrix. On the other hand, the IPN covariance matrix in \cite{zheng2019robustt} is reconstructed based on the interference power estimation. In \cite{zheng2019robust} improved beamformers based on reconstructing the interference-plus-noise covariance matrix were proposed, which can be used for the case of arbitrary steering vector mismatch, while the algorithm in \cite{chen2018adaptive}  reconstructs an IPN covariance matrix directly from the signal-interference subspace.
In \cite{MEPSalgorithm}, a method based on the principle of maximum entropy power spectrum (MEPS) is proposed for IPN matrix reconstruction in which the Capon spectrum estimation is replaced by the maximum entropy algorithm, while a reduced computational complexity version of this paper is presented in \cite{mohammadzadeh2020low}. The method proposed in \cite{zhu2020robust} utilizes the orthogonality of the noise and signal subspaces to reconstruct the IPN covariance matrix. In \cite{yang2021robust} the IPN matrix is reconstructed based on interference power estimation via the quasi-orthogonality between different steering vectors and the colored noise covariance matrix. In \cite{sun2021robust}, a method for reconstructing the IPN covariance matrix was developed based on gradient searching to obtain more accurate steering vectors,
while algorithm in \cite{mohammadzadeh2021robust} utilizes the virtual received steering vector and the spatial sampling process to reconstruct the IPN matrix based on an orthogonal projection matrix and retain the interference plus noise components and recently, the method proposed in \cite{mohammadzadeh2022robust} involves IPN matrix reconstruction using an idea in which the power of the interferences is estimated based on the power method.

Although the aforementioned IPN covariance matrix reconstruction approaches considerably enhance beamforming performance, it is worth noting that all of the above-mentioned methods can only generate narrow nulls in the interference incident direction. When the interference moves or the array platform shakes, the interference may move out of the null and cannot be effectively suppressed. Therefore, the ability of the beamformer to suppress interference is reduced, and may even completely fail.

\subsection{Contributions}

In this paper, we propose an efficient adaptive beamforming algorithm based on the IPN covariance matrix.  The proposed method, which we denote by IPN-PSEUR, relies on power spectral estimation and on the uncertainty region around the IPN components. To begin with, the initial direction of arrivals (DoAs) of the desired signal and interferences are estimated by the spectral MUltiple SIgnal Classification (MUSIC) approach. Then, instead of determining the pre-defined or fixed angular sector of interference, the proposed approach generates a procedure in every snapshot to adjust an uncertainty region for each DoA of interference with small angular sector. Moreover, to shape the directional response of the beamformer, a simplified power spectral density function based on a piecewise constant function with only two levels of the interference plus noise component is introduced, which results in a low-cost, but effective approach to IPN covariance matrix reconstruction. The proposed IPN-PSEUR beamformer can create notches in the directional response of the array with adequate depths and widths to effectively suppress interference signals.

The rest of this paper is arranged as follows. In section \ref{Sec II}, we introduce the signal model and provide background on adaptive beamforming. The proposed methods to estimate the IPN uncertainty region and IPN covariance matrix are presented in Section \ref{Sec III}. The mathematical analysis and computational complexity of the proposed PSEUR method are given in \ref{Analysis}, while the pseudo code is summarized in Algorithm 1. Simulation results are described in Section \ref{Simulation}, and Section \ref{Conclusion} concludes this paper.

\textit{Notation}: Lower case is used for scalar quantities (e.g., $v$) and bold lower case for column vectors (e.g., $\mathbf{v}$). Bold capital letters represent matrices (e.g., $\mathbf{C}$). $\mathbf{c}_y$ stands for the $y^{th}$ column of $\mathbf{C}$. For a vector $\mathbf{v}$, we denote its $y^{th}$ element as $\mathbf{v}_y$. $(\cdot)^T$ stands for transposition, while $\mathcal{E}\left\{\cdot \right\}$ and $ (\cdot)^H $ denote statistical expectation and Hermitian of a matrix or vector, respectively. The rank of a matrix is denoted by $\text{rank}(\cdot)$, the span of a matrix is denoted by $\text{span}(\cdot)$, while $\text{diag} (\cdot) $ defines a diagonal matrix. $\|\cdot\|_\mathrm{F}^2$ is the Frobenius norm, while $\|\cdot\|_2$ is the $\ell_2$ norm. $\mathbf{I}_M$ represents the $M \times M$ identity matrix.

\section{Data Model and Problem Statement} \label{Sec II}
Consider that there are $K$ far-field signal sources including a desired signal and $K-1$ interferers impinging on an $M$-element antenna array. The spacing between elements, $d$, is equal to one-half wavelength. The received data at time $t$, $\mathbf{x}(t)$, is modeled as
\begin{equation} \label{recieved vector}
\mathbf{x}(t)=\mathbf{a}(\theta_{1}) s_{1}(t)+\sum_{k=2}^{K} \mathbf{a}(\theta_{k}) s_{k}(t)+\mathbf{n}(t),
\end{equation}
where $\mathbf{x}(t)=[\begin{matrix}x_1(t) & \dots & x_{M}(t)\end{matrix}]^T$, and the signals $x_m(t)$, $m=1,2,\cdots, M$ are the output of each antenna element, while $ s_0(t) $ and $ s_k(t) $ denote the components of the SOI and $k^{th}$ interference, respectively; $\mathbf{n}(t)$ is an additive white Gaussian noise vector with zero mean and covariance matrix $\sigma^2_n \mathbf{I}_M$; $\mathbf{a}(\theta_{1})$ and $\mathbf{a}(\theta_{k})$ represent the $M \times 1$ steering vector from the desired signal direction $\theta_1$ and interference directions $\theta_k$, respectively. In this work, we assume that the SOI, interferences, and noise are statistically independent of each other. Consequently the covariance matrix of the received signal vector may be
expressed as
\begin{align} \label{Theoretical R}
\mathbf{R}_{xx} &=\mathcal{E}\left\{\mathbf{x}(t) \mathbf{x}^{{H}}(t)\right\}= \mathbf{A} \mathbf{R}_\mathrm{s} \mathbf{A}^H+ \sigma_{n}^{2} \mathbf{I}_M\nonumber \\
&=\underbrace{\sigma_1^{2} \mathbf{a}(\theta_1) \mathbf{a}^{H}(\theta_1)+ \sum_{k=2}^{K} \sigma_{k}^{2} \mathbf{a}(\theta_k) \mathbf{a}^{H}(\theta_k)}_{ = \mathbf{A} \mathbf{R}_\mathrm{s} \mathbf{A}^H} +\sigma_{n}^{2} \mathbf{I}_M,
\end{align}
where $\mathbf{R}_\mathrm{s}=\mathcal{E}\left\{\mathbf{s}(t)\mathbf{s}^H(t) \right\}$ is a covariance matrix with rank $K$ ($\text{rank}(\mathbf{A})=K$)
and $\mathbf{A}=[\mathbf{a}(\theta_1), \cdots,
\mathbf{a}(\theta_K)]$ is a Vandermonde matrix. Also, $\sigma^2_1$ and $\sigma^2_k$ denote the powers of the desired signal and the $k^{th}$ interference. At direction $\theta$, the steering vector $\mathbf{a}(\theta)$ is given by
\begin{equation}
    \mathbf{a}(\theta)=\big[1\ e^{-j (2 \pi d/ \lambda) \sin \theta} \cdots\ e^{-j  (M-1) (2 \pi d/ \lambda) \sin \theta} \big]^T,
\label{eq:a.def}\end{equation}
where $\lambda $ represents the signal wavelength and $d=\lambda/2 $ is the interelement spacing. \\
\indent It is well-known that the standard Capon beamformer (SCB) is a representative example of adaptive array beamformer which intends to allow the SOI to pass through without any distortion while the interference signals and noise are suppressed as much as possible, thereby maximizing the output signal-to-interference-plus-noise ratio (SINR). The SCB can be formulated as
\begin{align}\label{MVDR}
\underset{{\mathbf{w}}}{\operatorname{min}}\ \mathbf{w}^H \mathbf{R}_\mathrm{i+n} \ \mathbf{w}\ \hspace{.4cm} \mathrm{s.t.} \hspace{.4cm} \mathbf{w}^H \mathbf{a}(\theta_1)=1.
\end{align}
Here matrix
$\mathbf{R}_\mathrm{i+n}=\sum_{k=2}^{K} \sigma_{k}^{2} \mathbf{a}(\theta_k) \mathbf{a}^{H}(\theta_k)+\sigma_{n}^{2} \mathbf{I}_M$ denotes the theoretical IPN covariance matrix.
The solution to (\ref{MVDR}) yields the optimal beamformer given by
\begin{align}\label{optimal wegight vector}
\mathbf{w}_{\mathrm{opt}}=\dfrac{\mathbf{R}_\mathrm{i+n}^{-1} \mathbf{a}(\theta_1)}{\mathbf{a}(\theta_1)^H \mathbf{R}_\mathrm{i+n}^{-1}\mathbf{a}(\theta_1)},
\end{align}
where $\mathbf{w}=[w_1, \cdots, w_M]^T$ is the beamformer weight vector. The optimal beamformer weight vector $\mathbf{w}$ can be obtained by maximizing the output SINR as follows
\begin{align}\label{SINR}
\mathrm{SINR}=\dfrac{\sigma^{2}_\mathrm{1} |\mathbf{w}^H \mathbf{a}(\theta_1)|^2 }{\mathbf{w}^H \mathbf{R}_\mathrm{i+n}\mathbf{w}}.
\end{align}
In practice, the IPN covariance matrix,
$ \mathbf{R}_\mathrm{i+n} $ is replaced with the sample covariance matrix
\begin{align}
    \hat{\mathbf{R}}_{xx}=\dfrac{1}{N}\sum_{n=1}^{N} \mathbf{x}(t_n)\mathbf{x}^H(t_n),
\end{align}
where $N$ is the number of snapshots.

\section{The Proposed IPN-PSEUR Algorithm} \label{Sec III}

In this section, the spectral MUSIC method for DoA estimation is utilized and a novel beamforming algorithm, denoted IPN-PSEUR (the acronym refers to power spectrum and uncertainty  region) Q is introduced. Then, a new approach for  power spectral estimation of the IPN components will be addressed to simplify the process of the IPN covariance matrix reconstruction. The main idea of the proposed method is to  estimate the power and region where the interference signals lie, then obtain a more precise reconstructed IPN covariance matrix.

\subsection{Estimation of IPN Uncertainty Region}
One of the most important and critical problems facing sensor array systems is the detection of the number of sources impinging on the array \cite{spa}. This is a key step in most super-resolution estimation techniques \cite{jio,jidf,jiodoa,alrdoa,mskaesprit}, which often take the number of signals as prior information. However, in this work, we utilize a MUSIC-based criterion \cite{christensen2009sinusoidal} which provides asymptotically unbiased estimates of 1) the number of incident wavefronts present; 2) directions of arrival (or emitter locations).  It is assumed that the columns of $\mathbf{A}$ are linearly independent, the angles $\theta_k$ are distinct and $\mathbf{A}$ has rank $K$. Consider the eigendecomposition of $\mathbf{R}_{xx}$ as follows
\begin{equation} \label{EVD}
    \mathbf{R}_{xx}= \mathbf{U}_\mathbf{s} \mathbf{\Lambda}_\mathbf{s} \mathbf{U}_\mathbf{s}^H+ \mathbf{U}_\mathbf{n} \mathbf{\Lambda}_\mathbf{n} \mathbf{U}_\mathbf{n}^H,
\end{equation}
where $\mathbf{\Lambda}_\mathrm{s}=\text{diag}(\lambda_1,\dots,\lambda_K)$ and $\mathbf{\Lambda}_\mathrm{n}=\text{diag}(\lambda_{K+1},\dots,\lambda_M)$ are the eigenvalues of matrix $\mathbf{R}_{xx}$ corresponding to the signal subspace and noise subspace, respectively, which are in descending order $\lambda_1 \geq \cdots \geq \lambda_K \geq \lambda_{K+1}, \cdots, \lambda_M$, and the corresponding eigenvectors are $\mathbf{u}_1, \cdots, \mathbf{u}_K, \mathbf{u}_{K+1}, \cdots, \mathbf{u}_M$. $\mathbf{U}_\mathrm{s}$ is the eigenvector matrix composed of the eigenvectors associated with the first $K$ larger eigenvalues, which spans the signal subspace, and, $\mathbf{U}_\mathrm{n}$ is the eigenvector matrix composed of the eigenvectors associated with the first $M-K$ least significant eigenvalues which span the noise subspace where $\mathbf{U}_\mathrm{s}=[\mathbf{u}_1, \cdots, \mathbf{u}_K] \perp \mathbf{U}_\mathrm{n}=[\mathbf{u}_{K+1}, \cdots, \mathbf{u}_M]$. By comparing \eqref{EVD} and \eqref{Theoretical R}, we can see the columns of $\mathbf{A}$ span the same space as the columns of $\mathbf{U}_\mathrm{s}$, ($\text{span}(\mathbf{U}_\mathrm{s})=\text{span}(\mathbf{A})$) and $\text{span}(\mathbf{U}_\mathrm{n})$ is orthogonal to $\text{span}(\mathbf{U}_\mathrm{s})=\text{\text{span}}(\mathbf{A})$. Thereby, we can write
\begin{align} \label{Orth Sub}
    \mathbf{A}^H \mathbf{U}_\mathrm{n}= \mathbf{0}.
\end{align}
Then, the set of directions $\theta_k$ are found by minimizing the Frobenius norm of \eqref{Orth Sub} as follows
\begin{align}
    \lbrace \hat{\theta}_k \rbrace= \underset{{ \hat{\theta}_k }}{\operatorname{argmin}} \| \mathbf{A}^H \mathbf{U}_\mathrm{n} \|_\mathrm{F}^2.
\end{align}
Since the signal steering vectors are orthogonal to $\mathbf{U}_\mathrm{n}$, we can find the individual directions as
\begin{align} \label{thetahat_est}
    \hat{\theta}_k = \underset{{\hat{\theta}_k}}{\operatorname{argmin}} \| \mathbf{a}^H(\theta_k) \mathbf{U}_\mathrm{n} \|_\mathrm{F}^2, \quad  k=1, \cdots, K
\end{align}
where $\hat{\theta}_k$ are the estimated DoA of signals. However, the orthogonality property in \eqref{Orth Sub} only holds approximately since eigenvectors and the covariance matrix are estimated from a finite set of vectors.

To mitigate the uncertainties regarding the source direction, estimation of the DoAs of the interference must consider the corresponding uncertain regions during the time interval in which the snapshots are taken. The need for using broad nulls often arises when the direction of arrival of the unwanted interference may vary slightly with time or may not be known exactly. A sharp null would require continuous steering for obtaining a reasonable value for the SNR. In the proposed technique, a procedure is described by identifying an angular sector that allows the algorithm to adjust the null in the beampattern toward interferers. To this end, we consider the case where the interferers can move with respect to the array, such that their directions vary with time as
\begin{align} \label{thetan}
    \theta_k(t_n)=\hat{\theta}_k+\frac{\Delta \theta_k(t_n)}{N} (t_n- \frac{N}{2}), \ \ \ n=1, \cdots, N
\end{align}
where $k=2, \cdots, K$, $\Delta \theta_k(t_n)$ is the spatial uncertainty region change during the observation interval, and $\hat{\theta}_k$ is the direction at the center of the observation interval. It is assumed that, during the observation interval, the interference direction, as observed by the array, stays in the interval $(\hat{\theta}_k- \Delta \theta_k/2, \hat{\theta}_k + \Delta \theta_k/2)$. In order to estimate $\Delta\theta_k$, we use the DoA estimation technique in \cite{mohammadzadeh2019robust}. This technique is based on using the correlation between the inner products of the steering vector corresponding to a general direction of incidence with the received vectors. Regarding this method, the angle that maximizes the magnitude of the inner product is taken as the DoA estimate in \eqref{thetahat_est}  by scanning an angular sector centred on $\hat{\theta}_k$ as \cite{mohammadzadeh2019robust}
\begin{align} \label{correlation estimator}
\theta_k(t_n)=\underset{\theta \in \Theta_\mathrm{x}}{\operatorname{argmax}}\ | \mathbf{x}^H(t_n)\mathbf{a}(\theta)|, \quad n=1,\cdots,N
\end{align}
where $ \Theta_\mathrm{x}= [\hat{\theta}_k- c, \hat{\theta}_k+ c] $ is the angular sector corresponding to the estimated interference signal in \eqref{thetahat_est} while $c\ll \pi $ is a small angle. It is noted that the parameter $c$ should be chosen large enough to guarantee that the correct direction is in the interval, but its exact value is not very important. By comparing \eqref{thetan} and \eqref{correlation estimator} and rearranging we can find $\Delta \theta_k(t_n)$ for $n=1,\cdots, N$ as
\begin{align} \label{Uncertainty region}
    \hat{\theta}_k+\frac{\Delta \theta_k(t_n)}{N} (t_n- \frac{N}{2})= \underset{\theta \in \Theta_\mathrm{x}}{\operatorname{argmax}}\  |\mathbf{x}^H(t_n)\mathbf{a}(\theta)| \nonumber \\  \Delta \theta_k(t_n)= \big(\frac{2N}{2t_n-N} \big) \big(\underset{\theta \in \Theta_\mathrm{x}}{\operatorname{argmax}}\  | \mathbf{x}^H(t_n)\mathbf{a}(\theta) | - \hat{\theta}_k \big).
\end{align}

\subsection{IPN Covariance Matrix Reconstruction}
Recent works on robust adaptive beamforming focused on reconstructing the IPN covariance matrix utilizing the power spatial spectrum distribution over all possible direction, obtained using either Capon spectrum estimation \cite{gu2012robust} or maximum entropy \cite{MEPSalgorithm} approaches as follows
\begin{align}\label{Rin definition}
\hat{\mathbf{R}}_{\mathrm{in}}=\int_{\Theta_{\text{ipn}}} \hat{\rho}(\theta) \mathbf{{a}}(\theta)\mathbf{{a}}^H(\theta)d\theta,
\end{align}
where $\Theta_{\text{ipn}} $ is assumed to be the angular sector of the interferences, $\hat{\rho}(\theta)$ is the power spectrum in the IPN spatial domain
\begin{equation}
 \hat{\rho}(\theta)=\frac{1}{\mathbf{a}^H(\theta)\hat{\mathbf{R}}^{-1}\mathbf{a}(\theta)} \; \cite{gu2012robust},  \hat{\rho}(\theta)=  \dfrac{1}{\epsilon|\mathbf{a}^H(\theta)\hat{\mathbf{R}}^{-1} \mathbf{u}_1|^2},  \; \cite{MEPSalgorithm}
\end{equation}
and $\mathbf{u}_1=[\begin{smallmatrix}1 & 0 & \cdots & 0\end{smallmatrix}]^\mathrm{T}$,   $\epsilon=1/\mathbf{u}_1^\mathrm{T}\hat{\mathbf{R}}^{-1}\mathbf{u}_1$. 
To approximate the integral, it is assumed in \cite{MEPSalgorithm,chen2015robust,zheng2018covariance} that $\Theta_{\text{ipn}}$ is uniformly sampled with $P$ sample points. Therefore, $\hat{\mathbf{R}}_{\text{in}}$ is calculated by
\begin{align}\label{App Ripn}
\hat{\mathbf{R}}_{\text{in}}=\sum_{p=1}^{P} \hat{\rho}(\theta_p) \mathbf{a}(\theta_p)\mathbf{a}^H(\theta_p)\Delta\theta,
\end{align}
where $\Delta\theta=|\Theta_\text{ipn}|/P$ and $|\Theta_\text{ipn}|=\text{length of }\Theta_\text{ipn}
$.

The main problems with this IPN matrix construction stem from the approximation of the integral of the rank one matrices with a summation that requires a large number of computations to be able to accurately synthesize powers from signals for the whole $2 \pi$ sector \cite{mohammadzadeh2021robust}. Moreover, errors in interference power estimation and corresponding angular sector lead to inaccuracies to suppress the unwanted signals with deep nulls.

To address these problems, a simple and novel method is utilized based on a piecewise constant approximation to $\hat{\rho}(\theta)$ with only two levels and the use of the estimated uncertainty sectors for interferences  given by \eqref{Uncertainty region}.
\begin{figure}[!]
    \centering
    \includegraphics[width=2.1in,height=1.4in]{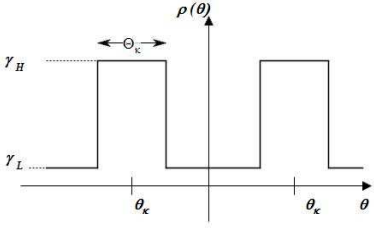}
\vspace{-0.75em}
    \caption{Power Spectrum of Uncertainty Region }
    \label{Exrapolation}
\end{figure}
As shown in Fig.~\ref{Exrapolation}, we concentrate most samples to small sectors around the interference angular sector. This is in contrast to other methods in which a large sector $[-\pi, \pi]$ is considered to find the interference directions. In particular, the parameter $ \hat{\rho}(\theta) $ is designed based on the fact that the strong interference signals should be suppressed more.
Thereby, we use an approximation for $\hat{\rho}(\theta)$ with just two levels, a low level $ \hat{\gamma}_\mathrm{L} $, and a high level $ \hat{\gamma}_\mathrm{H}$. It means that $\gamma_\mathrm{H} \gg \gamma_\mathrm{L}$. Therefore, we can modify the IPN covariance matrix in \eqref{Rin definition} as follows
\begin{align} \label{Rhatint}
\hat{\mathbf{R}}_\mathrm{in}=\hat{\gamma}_\mathrm{L} \int_{-\pi}^{\pi} &\mathbf{a}(\theta) \mathbf{a}^{H}(\theta) d \theta \nonumber \\
&+(\hat{\gamma}_\mathrm{H}-\hat{\gamma}_\mathrm{L}) \int_{-\pi}^{\pi} I_{\Theta_k}(\theta) \mathbf{a}(\theta) \mathbf{a}^{H}(\theta) d \theta,
\end{align}
where
\begin{equation}
I_{\Theta_k}(\theta_k)=\left\{\begin{array}{ll}
1, & \theta_k \in \Theta_k \\
0, & \text{Otherwise}
\end{array}\right.
\end{equation}
and $\Theta_k=(\hat{\theta}_k- \Delta \theta_k/2, \hat{\theta}_k + \Delta \theta_k/2)$.
In order to calculate the first integral, we utilize the definition of array steering vector \eqref{eq:a.def}.  Letting $u = u(\theta)=(2\pi d/\lambda)\sin\theta$
\begin{align}
\mathbf{a}(\theta) \mathbf{a}^{H}(\theta)&=
    \begin{bmatrix}
1&e^{j u}&\cdots& e^{j (M-1) u}\\
e^{-j u}& 1&\cdots& e^{j (M-2)u}\\
\vdots&\vdots&     &\vdots&\\
e^{-j (M-1) u}& e^{-j (M-2) u}& \cdots & 1
\end{bmatrix} \nonumber \\ &= [e^{j (\ell-m) u(\theta)}]_{\ell,m}, \ \ell,m=1,\cdots, M.
\end{align}
It can be shown that
\begin{align} \label{in aaH}
    \int_{-\pi}^{\pi} \mathbf{a}(\theta) \mathbf{a}^{H}(\theta) d \theta=2 \pi\left[ \mathbf{B} \right]_{\ell, m}.
\end{align}
where $\left[ \mathbf{B} \right]_{\ell, m}= \mathbf{B}_0 (|\ell-m| \pi)$ is the Bessel function of the first kind. Since $\left[ \mathbf{B} \right]_{\ell = m}=\text{diag}(\mathbf{B})=\mathbf{I}$, we can write
\begin{align} \label{integral aaH}
    \int_{-\pi}^{\pi} \mathbf{a}(\theta) \mathbf{a}^{H}(\theta) d \theta=2 \pi \mathbf{I}+ 2 \pi\left[ \mathbf{B}\right]_{\ell \neq m} \nonumber \\
   \hat{\gamma}_\mathrm{L} \int_{-\pi}^{\pi} \mathbf{a}(\theta) \mathbf{a}^{H}(\theta) d \theta=2 \pi \hat{\gamma}_\mathrm{L} \mathbf{I} + 2 \pi \hat{\gamma}_\mathrm{L} \mathbf{C}
\end{align}
where $\mathbf{C}= \left[ \mathbf{B} \right]_{\ell \neq m}$. Using \eqref{integral aaH},  we can re-write \eqref{Rhatint} as
\begin{align} \label{Rin with Integral}
\hat{\mathbf{R}}_\mathrm{in}&= 2 \pi\hat{\gamma}_\mathrm{L} \mathbf{I}+2 \pi \hat{\gamma}_\mathrm{L} \mathbf{C}+(\hat{\gamma}_\mathrm{H}-\hat{\gamma}_\mathrm{L}) \int_{\Theta_\mathrm{in}} \mathbf{a}(\theta) \mathbf{a}^{H}(\theta) d \theta \nonumber \\
&\approx 2 \pi \hat{\gamma}_\mathrm{L} \mathbf{I}+ (\hat{\gamma}_\mathrm{H}-\hat{\gamma}_\mathrm{L}) \int_{\Theta_\mathrm{in}} \mathbf{a}(\theta) \mathbf{a}^{H}(\theta) d \theta,
\end{align}
where $\Theta_\mathrm{in}=\bigcup_{k=2}^{K} \Theta_k$ and the approximation holds for $\hat{\gamma}_\mathrm{H}\gg\hat{\gamma}_\mathrm{L}$. By sampling $\Theta_\mathrm{in}$ uniformly with $Q_\mathrm{in} \ll P$  sampling points, \eqref{Rin with Integral} can be approximated by
\begin{align} \label{New Rin}
\hat{\mathbf{R}}_\mathrm{in} & \cong 2 \pi \hat{\gamma}_\mathrm{L}  \mathbf{I}+(\hat{\gamma}_\mathrm{H}-\hat{\gamma}_\mathrm{L}) \sum_{i=1}^{Q_\mathrm{in}} \mathbf{a}(\theta_{n_i})  \mathbf{a}^{H}(\theta_{n_i})\Delta\theta  \nonumber \\
& =2 \pi \hat{\gamma}_\mathrm{L}  \mathbf{I}+(\hat{\gamma}_\mathrm{H}-\hat{\gamma}_\mathrm{L}) \mathbf{R}_\mathrm{in}^r  \nonumber \\ &= 2\pi\hat{\gamma}_\mathrm{L} ( \mathbf{I}+ \frac{\hat{\gamma}_\mathrm{H}-\hat{\gamma}_\mathrm{L}}{2 \pi \hat{\gamma}_\mathrm{L}}\mathbf{R}_\mathrm{in}^r),
\end{align}
where we assumed that $Q_\text{in}=|\Theta_\text{in}|/\Delta\theta$,  $\mathbf{R}_\mathrm{in}^r=\sum_{i=1}^{Q_\mathrm{in}} \mathbf{a}(\theta_{n_i})  \mathbf{a}^{H}(\theta_{n_i}) \Delta\theta$.  Let the eigenvalue decomposition (EVD) of $ \mathbf{R}_\mathrm{in}^r $ be written as
\begin{align}\label{R_r}
\mathbf{R}_\mathrm{in}^r =\mathbf{E} \mathbf{\Xi} \mathbf{E}^H=\sum_{r=1}^R \xi_r \mathbf{e}_r \mathbf{e}^H_r,
\end{align}
where $ \mathbf{\Xi} $ is a diagonal matrix with the eigenvalues $\xi_r(\mathbf{R}_\mathrm{in}^r)(r=1, \cdots, R)$ of $\mathbf{R}_\mathrm{in}^r$  on the diagonal, and $\mathbf{E}$ is an orthogonal matrix with a corresponding set of orthonormal eigenvectors $\mathbf{e}_1, \cdots, \mathbf{e}_R$ as columns and $ (\text{rank}(\mathbf{R}_\mathrm{in}^r)=R \leq M) $ denotes the rank of $ \mathbf{R}_\mathrm{in}^r $. The rank $R$ depends on the width of the angular sector $ \Theta_k $, and is one if the width shrinks to zero. But the dominant eigenvalue would be of the order of $ \xi_\text{max}\approx M|\Theta_k| $ and the majority of the eigenvalues would be close to zero for a sufficiently small width or $|\Theta_k| \to 0$. By replacing (\ref{R_r}) back into \eqref{New Rin} and taking the inverse of $ \hat{\mathbf{R}}_\mathrm{in} $ based on the Woodbury matrix inversion lemma, we can show that
\begin{align}\label{C inv}
\hat{\mathbf{R}}_\mathrm{in}^{-1}=\frac{1}{2 \pi \hat{\gamma}_\mathrm{L}}\Big[\mathbf{{I}}_M-\mathbf{{E}}\Big( \frac{2 \pi \hat{\gamma}_\mathrm{L}}{ \hat{\gamma}_{H}- \hat{\gamma}_\mathrm{L}} \mathbf{\Xi}^{-1}+\mathbf{{E}}^H\mathbf{{E}} \Big)^{-1} \mathbf{{E}}^H \Big]\!\! \raisetag{17pt}
\end{align}
Also, based on the theoretical definition of the IPN covariance matrix, we have
\begin{equation} \label{Theoretical Rin}
   \mathbf{R}_\mathrm{i+n} =\sum_{k=2}^{K} \sigma_{k}^{2} \mathbf{a}(\theta_k) \mathbf{a}^{\mathrm{H}}(\theta_k)+\sigma_{n}^{2} \mathbf{I}_M = \mathbf{A}_K \mathbf{\Sigma}_K \mathbf{A}_K^H+ \sigma^2_n \mathbf{I}
\end{equation}
Then, the inverse of the IPN covariance matrix in (\ref{Theoretical Rin})
by applying  Woodbury's matrix inversion lemma is expressed as
\begin{align}\label{R theo inv}
\mathbf{R}_\mathrm{i+n}^{-1}=\frac{1}{\sigma^2_n}\Big[\mathbf{{I}}_M-\mathbf{A}_K\Big( \sigma^2_n \mathbf{\Sigma}_K^{-1}+\mathbf{A}_K^H\mathbf{A}_K \Big)^{-1} \mathbf{A}_K^H \Big],
\end{align}
where $\mathbf{A}_K=[\mathbf{a}(\theta_2), \cdots, \mathbf{a}(\theta_K)]$ and  $\mathbf{\Sigma}_K$ is the diagonal matrix with eigenvalues $\sigma_k^2 (\mathbf{R}_\mathrm{i+n})$ for $k=2, \cdots, K$.

By comparing \eqref{C inv} and \eqref{R theo inv}, we can easily see that
\begin{align}\label{eq:gammaL}
    \frac{1}{2 \pi \hat{\gamma}_\mathrm{L}} \cong \frac{1}{\hat{\sigma}^2_n} \quad \text{and thus} \quad   \hat{\gamma}_\mathrm{L}=\frac{1}{2 \pi} \hat{\sigma}^2_n,
\end{align}
where $\hat{\sigma}^2_n=\frac{1}{M-K} \sum_{i=K+1}^M \lambda^2_i$.
On the other hand, in order to effectively suppress interference, the choice of $\hat{\gamma}_\mathrm{H}$ is based on the fact that the interference with higher power should be more suppressed. To this end, the condition $\hat{\gamma}_\mathrm{H} \geq \text{max} \lbrace \hat{\sigma}_k^2 \rbrace_{k=2}^K $ is considered. Therefore, we can write
\begin{equation}
    \frac{2 \pi \hat{\gamma}_\mathrm{L}}{ \hat{\gamma}_\mathrm{H}- \hat{\gamma}_\mathrm{L}} \cong \frac{\hat{\sigma}^2_n}{\text{max} \lbrace \hat{\sigma}_k^2 \rbrace_{k=2}^K} \longrightarrow \hat{\gamma}_\mathrm{H}=\text{max} \lbrace \hat{\sigma}_k^2 \rbrace_{k=2}^K+ \hat{\gamma}_\mathrm{L}.\!\! \raisetag{17pt}
\end{equation}
\indent In order to find $\hat{\gamma}_\mathrm{H}$, the interference power is estimated via the quasi-orthogonality between different steering vectors for the sample covariance matrix. Let us recall the received data \eqref{recieved vector} as
\begin{equation}
\mathbf{x}(t)=\mathbf{a}_1 s_{1}(t)+\sum_{k=2}^{K} \mathbf{a}_k s_{k}(t)+\mathbf{n}(t).
\end{equation}
For brevity, we assume that $\mathbf{a}(\theta_{k})=\mathbf{a}_k$ and $\mathbf{a}(\theta_{1})=\mathbf{a}_1$. Pre-multiplying the above equation by  $\mathbf{a}^H_k$, we have
\begin{equation}
\mathbf{a}^H_k\mathbf{x}(t)=\mathbf{a}_k^H \mathbf{a}_1 s_{1}(t)+ \mathbf{a}_k^H \Big(\sum_{k=2}^{K} \mathbf{a}_k s_{k}(t)+\mathbf{n}(t) \Big).
\end{equation}
Assuming that $\mathbf{a}_{k}$ is approximately uncorrelated with $\mathbf{a}_j$ for $j\neq k$ (which would be true if $M$ is large enough and the angles $\theta_j$ and $\theta_k$ are sufficiently spaced for $j\neq k$) and the inner products of $\mathbf{a}_j$ and $\mathbf{a}_k $ are negligibly small for $k=2,\cdots,K$, we obtain
\begin{equation} \label{x_k}
z_k(t)\approx\mathbf{a}_k^H \mathbf{a}_k s_{k}(t)+ \mathbf{a}_k^H\mathbf{n}(t),
\end{equation}
so the sample variance can be written as
\begin{equation} \label{Rxk}
     \hat{r}_{z_k}=\frac{1}{N}\sum_{n=1}^{N} |z_k(t_n)|^2= \hat{\sigma}^2_k \mathbf{a}_k^H \mathbf{a}_k \mathbf{a}_k^H \mathbf{a}_k + \hat{\sigma}_n^2 \mathbf{a}_k^H \mathbf{a}_k,
\end{equation}
where $\hat{\sigma}_k^2$ is an approximation to $\sigma_k^2=\mathcal{E}\left\{ |s_k(t) |^2 \right\}$, and $\hat{\sigma}_n^2$ is given after \eqref{eq:gammaL}. The power of the $k$-th interference $\hat{\sigma}_k^2$ can be obtained, recalling that $\mathbf{a}_k^H\mathbf{a}_k=M$, by
\begin{align} \label{sigmak}
    \hat{\sigma}_k^2=\frac{\hat{r}_{z_k}-\hat{\sigma}_n^2 M}{M^2}.
\end{align}
Finally, for the assumed SOI direction $\hat{\theta}_1$ ($\mathbf{a}(\hat{\theta}_{1})=\hat{\mathbf{a}}_1$), the robust beamformer is computed by
\begin{align}\label{pseur}
\mathbf{w}_{\mathrm{PSEUR}}=\dfrac{\hat{\mathbf{R}}_\mathrm{in}^{-1} \mathbf{a}(\hat{\theta}_1)}{\mathbf{a}(\hat{\theta}_1)^\mathrm{H} \hat{\mathbf{R}}_\mathrm{in}^{-1}\mathbf{a}(\hat{\theta}_1)}.
\end{align}
\section{Analysis of the Proposed IPN-PSEUR Method} \label{Analysis}
In this section, we provide a mathematical analysis of the proposed IPN-PSEUR approach. We also compare the computational complexity of the proposed method and existing methods. A summary of the proposed IPN-PSEUR method is presented in Algorithm 1.

\subsection{Computational Complexity and Summary of IPN-PSEUR}

\begin{figure}[h]
    \centering
    \includegraphics[height=2.4in]{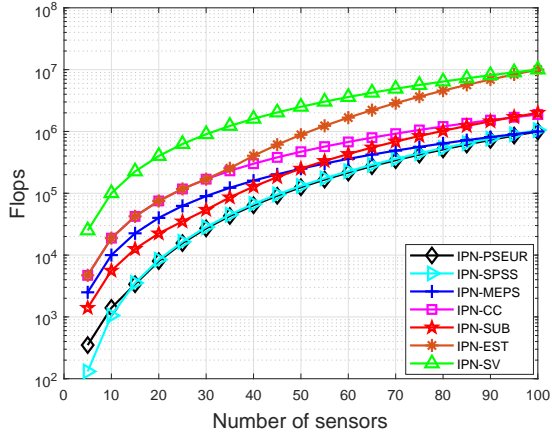}
    \caption{Growth rate for the computational complexity for the various algorithms}
    \label{Complexity}
\end{figure}

The reconstructed IPN-CC covariance matrix method in [12] has a complexity of $ \mathcal{O}(PM^2)$, while the IPN-SPSS [16] incurs a complexity of $ \mathcal{O}(M^3)$ because of the matrix inversion. The complexity of the  IPN-EST [18] and IPN-SV [33] is $\text{max}( \mathcal{O}(PM^2), \mathcal{O}(M^{3.5}))$ and $\text{max}( \mathcal{O}(P_i M^2), \mathcal{O}(M^{3.5}))$ respectively, since both need to solve a convex optimization problem with high computational cost. The IPN-SUB beamformer in [25] has computational complexity of $ K \times \text{max} \big( \mathcal{O}(P_l M^2),\mathcal{O}(PM^2 ) \big)$ where $P_l$ denotes the number of samples in the small angular sector of the interference region. In [23] (IPN-MEPS), the IPN covariance matrix is reconstructed with a complexity of $ \mathcal{O}(P_m M^2)$ and the desired signal steering vector estimation requires the complexity of $ \mathcal{O}(SM^2)$ where $S$ denotes the number of samples in the desired signal region. In all simulation results, the parameters $P=188$, $P_i=1002$, $P_m=90$, $S=14$, $Q_\text{in}=14$ and $P_l=28$ are considered. Note that, for each algorithm in the simulations, we employed the lowest quantities that result in the best performance because some references did not provide values for the parameter $P$.
\begin{table}[h] \label{com}
  \centering
  \caption{Computational time}
  \label{tab:expcond}
  \begin{tabular}{|c |c |}
    \hline
    Beamformers  &Execution time (s)    \\
    \hline \hline
IPN-SPSS   &$ 0.0065 $   \\ \hline
IPN-PSEUR  & $ 0.0103 $     \\ \hline
IPN-SUB   &$ 0.0222 $   \\ \hline
IPN-MEPS  &$ 0.0247 $   \\ \hline
IPN-CC   & $0.0301$     \\\hline
IPN-EST   &$ 0.7116 $    \\ \hline
IPN-SV  &$ 0.7669$    \\
    \hline
\end{tabular}
\end{table}
In Fig.~\ref{Complexity}, we compare the growth rate of the computational complexity  for the proposed algorithm and the other tested algorithms in terms of the number of sensors $M$ used for each snapshot, and assuming the other parameters as described above. The reconstruction of the INC matrix requires solving a quadratically-constrained quadratic program (QCQP), which increases the complexity of IPN-SV and IPN-EST. It is evident that the complexities of IPN-CC, IPN-MEPS, and IPN-SUB have almost the same slope against the number of sensors, while the IPN-SUB has low complexity since it allows a  nonuniform sampling for the interference-plus-noise angular sector. On the other hand, although the IPN-SPSS requires fewer computations compared to the other methods (including the proposed method), its performance of is not as good as that of the other reconstructed IPN algorithms. Note also that the IPN-SV method has a highly-variant computational complexity in different snapshots, due to the online optimization program used for the steering vector estimation. CMR-SV, on the other hand, needs more sampling points ($P_i=1002$) for the interference plus noise angular sector. Therefore, its computational complexity is high.\\
As an example to show the computational complexity of the proposed method and other compared methods, we ran all algorithms using MATLAB 2017a on a Windows 10 laptop with dual core 1.9 GHz Intel Core i3 CPU and 3.36 GB memory. the results are demonstrated in Table~\ref{tab:expcond}. To do this properly, we closed all background processes in our computer (internet, e-mail, music, all unrelated software), and restricted MATLAB to use a single processor all time. Furthermore, it should be noted that these measurements have been done based on the parameters given in the simulation section and the scenario of that there is no mismatch between the actual and assumed steering vector and exact signal steering vector is known (first scenario in the simulation section).

\begin{algorithm} [t]
    \caption{Proposed PSEUR Adaptive Beamforming }\label{sobelcode}
    1: \textbf{Input:}\:Array received data vector $\lbrace \mathbf{{x}}(t_n) \rbrace_{n=1}^N $,\\
    2:  Compute   ${\hspace{1em}}\hat{\mathbf{{R}}}_{xx}=(1/N\sum_{n=1}^{N} \mathbf{{x}}(t_n)\mathbf{{x}}^H(t_n)$;\\
    3: Construct  $\hat{\mathbf{R}}_{xx}=\mathbf{U} \mathbf{\Lambda} \mathbf{U}^H$, $\mathbf{S}=\mathbf{A}=[\mathbf{u}_1, \cdots, \mathbf{u}_K]$, $\mathbf{G}=[\mathbf{u}_{K+1}, \cdots, \mathbf{u}_M]$;   \\
    4: Compute $\hat{\sigma}^2_n=\frac{1}{M-K} \sum_{i=K+1}^M \lambda^2_i$ and $\hat{\gamma}_\mathrm{L}=\frac{1}{2 \pi} \hat{\sigma}^2_n $\\
    5: \textbf{For} $ k=2:K$ \\
    6:\quad  \ Compute $\hat{\theta}_k=\underset{{\lbrace \hat{\theta}_k \rbrace}}{\operatorname{argmin}} \| \mathbf{A}^H \mathbf{G} \|_\mathrm{F}^2$;\\
    7: \quad Calculate $\mathbf{a}_k=\mathbf{a}(\hat{\theta}_k)=[1, e^{-j \hat{\theta}_k}, \cdots, e^{-j  (M-1)\hat{\theta}_k} ]^T$\\
    8:  \quad \quad \textbf{For} $n=1:N$ \\
    9: \quad \qquad  Obtain $\Delta \theta_k(t_n)$ using \eqref{Uncertainty region}; \\
    10:\quad \qquad Compute $\lbrace z_k(t_n) \rbrace_{n=1}^N $ using \eqref{x_k};\\
    11:\quad \qquad Compute $\hat{r}_{z_k}$ using \eqref{Rxk};\\
    12:\quad \qquad Compute $\hat{\sigma}_k^2$ using \eqref{sigmak};\\
    13:\quad \quad \textbf{End}\\
    14: \quad Assign $\Theta_k=(\hat{\theta}_k- \Delta \theta_k/2, \hat{\theta}_k + \Delta \theta_k/2)$;\\
    15: \textbf{End} \\
    16: Calculate $\hat{\gamma}_\mathrm{H}=\text{max} \lbrace \hat{\sigma}_k^2 \rbrace_{k=2}^K+ \hat{\gamma}_\mathrm{L}$;\\
    17: Assign $\Theta_\mathrm{in}=\bigcup_{k=2}^{K} \Theta_k$, sampling $\Theta_\mathrm{in}$ uniformly with $Q_\mathrm{in}$  sampling points;\\
    18: Compute $\mathbf{R}_\mathrm{in}^r=\sum_{i=1}^{Q_\mathrm{in}} \mathbf{a}(\theta_{n_i})  \mathbf{a}^{H}(\theta_{n_i}) \Delta \theta_{n_i}$;\\
    19: Reconstruct IPN Covariance matrix, $\hat{\mathbf{R}}_\mathrm{in}$ by \eqref{New Rin};\\
    20 Design proposed beamformer using \eqref{pseur};\\
    21: \textbf{Output:}\: Proposed beamforming weight vector $ \mathbf{w}_{\mathrm{PSEUR}} $.
\end{algorithm}

\subsection{Theoretical Considerations on the Beampattern Within a Notch for the Proposed PSEUR Method}
This subsection derives the expression of the directional response of the array in the angular sector of interferences to investigate the performance of the proposed IPN-PSEUR approach. To this end, we employ the example of a single interference. Understanding the relationship between the depth of the notch in the beampattern and the weight vector is useful for the presented analysis.

A standard tool for analyzing the performance of a beamformer is the response for a proposed weight vector $\mathbf{w}_{\mathrm{PSEUR}}$ as a function of angle $\theta$, known as the beam response. This angular response is computed by applying the beamformer $\mathbf{w}_{\mathrm{PSEUR}}$ to a set of array response vectors from all possible angles, that is, $-90^o \leq \theta \leq 90^o$, as follows
\begin{align}\label{beampattern}
\mathbf{D}(\theta)=\big \rvert \mathbf{w}_{\mathrm{PSEUR}}^H \mathbf{a}(\theta) \big \rvert= \Big \rvert \dfrac{\mathbf{a}^{H}(\theta) \hat{\mathbf{R}}_{\mathrm{in}}^{-1} \hat{\mathbf{a}}_1}{\hat{\mathbf{a}}_1^{H} \hat{\mathbf{R}}_{\mathrm{in}}^{-1} \hat{\mathbf{a}}_1} \Big \rvert. \end{align}
The numerator of \eqref{beampattern} is computed by substituting (\ref{C inv}) into (\ref{beampattern}) as follows
\begin{align}\label{num}
\Big \rvert \mathbf{a}^H(\theta) &\hat{\mathbf{R}}_{\mathrm{in}}^{-1} \hat{\mathbf{a}}_1 \Big \rvert=\dfrac{1}{2\pi \hat{\gamma}_\mathrm{L}}\Big\rvert\mathbf{a}^{H}(\theta)\hat{\mathbf{a}}_1 \nonumber \\
&-\mathbf{a}^H(\theta)\mathbf{E}\Big( \frac{2 \pi \hat{\gamma}_\mathrm{L}}{ \hat{\gamma}_{H}- \hat{\gamma}_\mathrm{L}} \mathbf{\Xi}^{-1}+\mathbf{I}_R\Big)^{-1} \mathbf{E}^{H} \hat{\mathbf{a}}_1 \Big\rvert,
\end{align}
since $\mathbf{E}$ is an orthogonal matrix with a corresponding set of orthonormal eigenvectors $\mathbf{e}_1, \cdots, \mathbf{e}_R$ of $\mathbf{R}_\mathrm{in}^r$ as columns, we employ the fact that $ \mathbf{E}^{H}\mathbf{E}=\mathbf{\mathbf{I}}_R $. That means any steering vector whose DoA comes from $\Theta_k$ can be expressed as a linear combination of the columns of $\mathbf{E}$ \cite{lie2011adaptive}. Therefore, the steering vector $ \mathbf{a}(\theta) \in \text{span}(\mathbf{E}) $  can be expressed as $ \mathbf{{a}}(\theta)=\mathbf{E} \mathbf{h}(\theta) $ for some $\mathbf{h}(\theta)$ (and for $\theta\in\Theta_k$) {\cite{zhang2013robust}}.

In order to simplify the expression in (\ref{num}), we define a vector for $\theta\in\Theta_k$ which is the orthogonal projection of $ \hat{\mathbf{a}}_1 $ onto the subspace spanned by the eigenvectors $ \mathbf{e}_r$ for ($r=1,\cdots ,R$) as follows
\begin{align}
\mathbf{{E}}^{H} \hat{\mathbf{a}}_1=\mathbf{b}=[b_{1}, \cdots, b_R]^T,
\end{align}
then, \eqref{num} can be written as
\begin{align}
\Big \rvert \mathbf{a}^H(\theta) \hat{\mathbf{R}}_{\mathrm{in}}^{-1} \hat{\mathbf{a}}_1 \Big \rvert&=\Big\rvert \dfrac{  \mathbf{h}^{H}(\theta) }{2\pi \hat{\gamma}_\mathrm{L}}  \Big( \mathbf{b}
-\Big( \frac{2 \pi \hat{\gamma}_\mathrm{L}}{ \hat{\gamma}_{H}- \hat{\gamma}_\mathrm{L}}\mathbf{\Xi}^{-1}+\mathbf{I}_R\Big)^{-1} {\mathbf{b}} \Big) \Big\rvert \nonumber\\
&=\Big\rvert \sum_{r=1}^{R}h^*_r(\theta)b_r \Big(\dfrac{1}{2\pi \hat{\gamma}_\mathrm{L}+\xi_r (\hat{\gamma}_{H}- \hat{\gamma}_\mathrm{L})}\Big) \Big\rvert.
\end{align}
The denominator of (\ref{beampattern}) can be alternatively expressed by
\begin{align}\label{den}
\Big\rvert \hat{\mathbf{a}}_1^{H} \hat{\mathbf{R}}_{\mathrm{in}}^{-1} \hat{\mathbf{a}}_1 \Big \rvert&=\dfrac{1}{2\pi \hat{\gamma}_\mathrm{L}} \Big[\| \hat{\mathbf{a}}_1 \| ^2 \nonumber \\  & -\sum_{r=1}^{R} \vert b_r \vert^2 \Big(\dfrac{\xi_r (\hat{\gamma}_{H}- \hat{\gamma}_\mathrm{L})}{2\pi \hat{\gamma}_\mathrm{L}+\xi_r (\hat{\gamma}_{H}- \hat{\gamma}_\mathrm{L})}\Big) \Big].
\end{align}
If the SOI angular direction is sufficiently separated from the sector $ \Theta_k $, then $\| \mathbf{b} \|_2 << \|\hat{\mathbf{a}}_1 \|_2$.
Moreover, for a sufficiently small sector $ \Theta_k $, the dominant eigenvalue would be much bigger than most of the eigenvalues $ \lambda_r $ of $ \mathbf{R}_r $ which would be either very small or almost zero. According to these points, the summation in (\ref{den}) can be disregarded and the beampattern is expressed as follows
\begin{align} \label{last beampattern}
\mathbf{D}(\theta)=\dfrac{2\pi \hat{\gamma}_\mathrm{L}}{\| \hat{\mathbf{a}}_1 \| ^2 } \Big\vert \sum_{r=1}^{R}h^*_r(\theta)b_r \Big(\dfrac{1}{2\pi\hat{\gamma}_\mathrm{L}+\xi_r (\hat{\gamma}_{H}- \hat{\gamma}_\mathrm{L})}\Big) \Big\vert. \!\! \raisetag{17pt}
\end{align}
\eqref{last beampattern} illustrates that by choosing large values of $ (\hat{\gamma}_{H}- \hat{\gamma}_\mathrm{L}) $, the beampattern inside the notch would be small. Moreover, \eqref{den} implies that the parameter $ (\hat{\gamma}_{H}- \hat{\gamma}_\mathrm{L}) $ affects the array gain at directions outside the interference sector and the gain at these directions would be increased unnecessarily. Therefore, in order to eliminate the interference precisely, the value of $\hat{\gamma}_H$ should not be too large, only enough to suppress the interference.

\section{Simulations} \label{Simulation}
In the simulations presented in this paper, a ULA consisting of $ M=20 $ omnidirectional sensors is considered. In each sensor, the additive noise is modeled as a complex Gaussian process with zero mean and unit variance that is spatially and temporally white. The desired signal and two interfering sources are impinging from $\hat{\theta}_1=10^\circ $, $-50^\circ$ and $30^\circ$ respectively.  In all scenarios, we assume that the number of snapshots is $K=30$, the input SNR is fixed at 10 dB, and perform 100 Monte-Carlo runs while the interference to noise ratios (INRs) is set to 30 dB for both interferers.

The proposed (IPN-PSEUR) beamformer is compared with the following approaches: the subspace-based beamformer \cite{zhu2020robust} (IPN-SUB), the beamformer in \cite{chen2015robust} (IPN-CC), the reconstruction-estimation based beamformer \cite{zheng2018covariance} (IPN-EST), the beamformer in \cite{zhang2016interference} (IPN-SPSS), the beamformer in \cite{khabbazibasmenj2012robust} (IPN-SV) and the beamformer in \cite{MEPSalgorithm} (IPN-MEPS). The angular sector $\Theta_\mathrm{x}$ for the proposed method in \eqref{correlation estimator} is defined as $[{\hat{\theta}_k}-3^\circ,{\hat{\theta}_k}+3^\circ]$, and the angular sector of the desired signal is set to be $ {\Theta}_s=[{\hat{\theta}_1}-6^\circ,{\hat{\theta}_1}+6^\circ] $ where the interference angular sector is $ \Theta_\text{ipn}=[-90^\circ,{\hat{\theta}_1}-6^\circ)\cup({\hat{\theta}_1}+6^\circ,90^\circ] $ and the bound for the beamformer in \cite{zheng2018covariance} is set as $ \epsilon=\sqrt{0.1} $. The degree interval is $0.9^o$ in the simulations. In \cite{zhang2016interference} (IPN-SPSS), the reference angle is $\alpha_0=0^\circ$ and the null broadening parameter is $\Delta=\sin^{-1}(2/M)$. The energy percentage $\rho$ is set to 0.9 in IPN-SUB. The CVX toolbox is used to solve the optimization problem in other methods \cite{grant2008cvx}.

\subsection{Example 1: Exact signal steering vector is known}
\begin{figure}[h]
    \centering
    \includegraphics[height=2.3in]{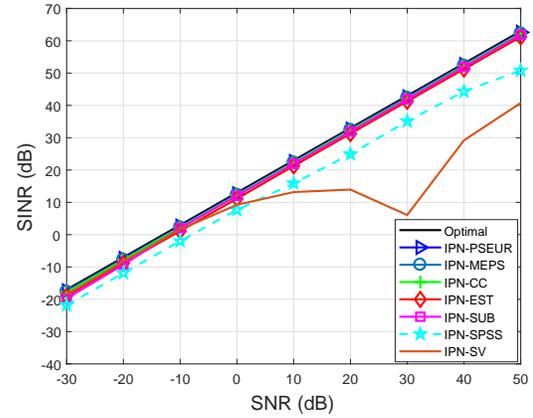}
\vspace{-0.75em}
    \caption{Output SINR versus SNR for Example 1}
    \label{SINR_SNR_No_Mismatch}
\end{figure}

\begin{figure}[h]
    \centering
    \includegraphics[height=2.3in]{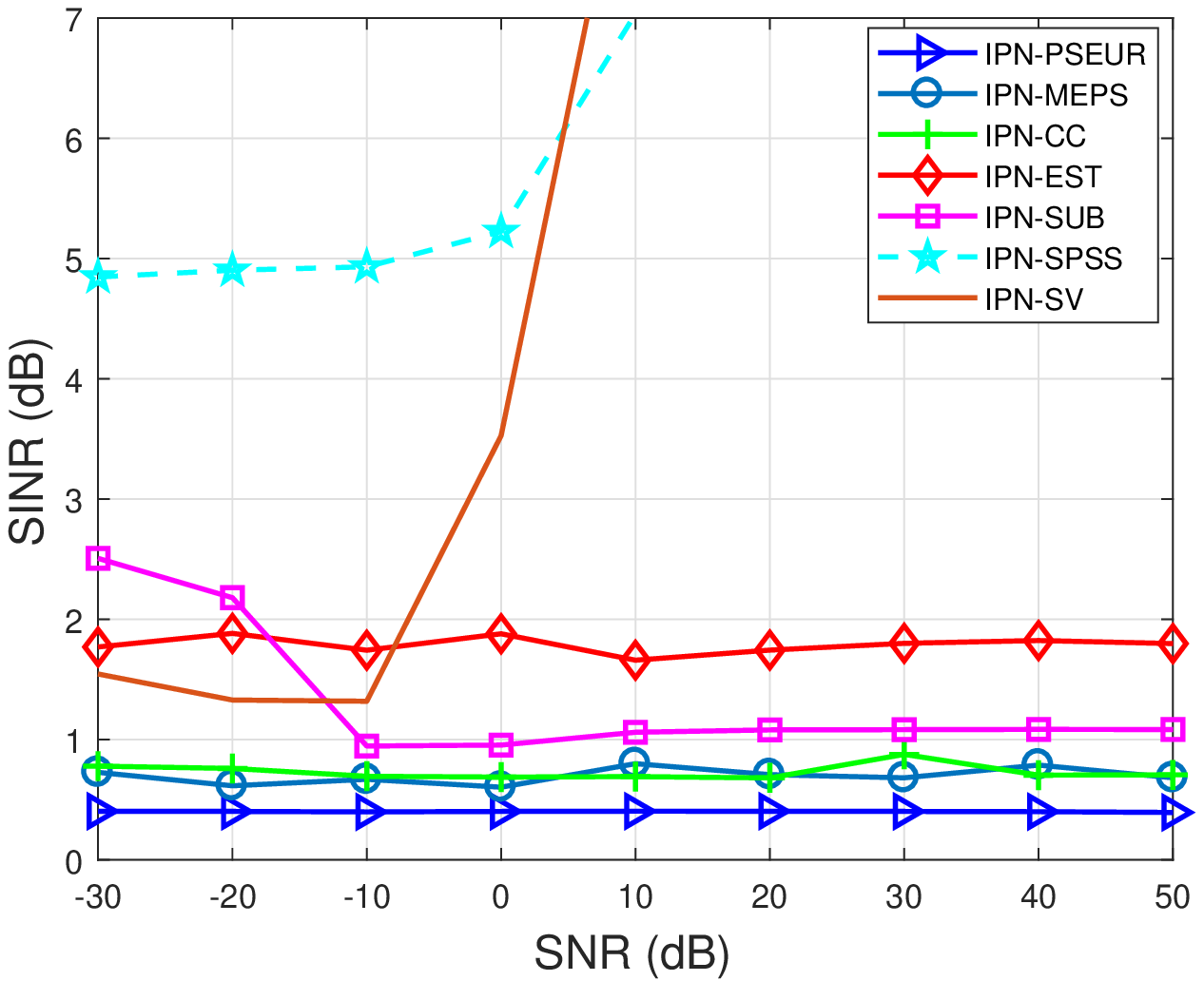}
\vspace{-0.75em}
    \caption{ Deviation from optimal SINR versus SNR for Example 1 }
    \label{SINR_Dev_No_Mismatch}
\end{figure}

\begin{figure}[h]
    \centering
    \includegraphics[height=2.3in]{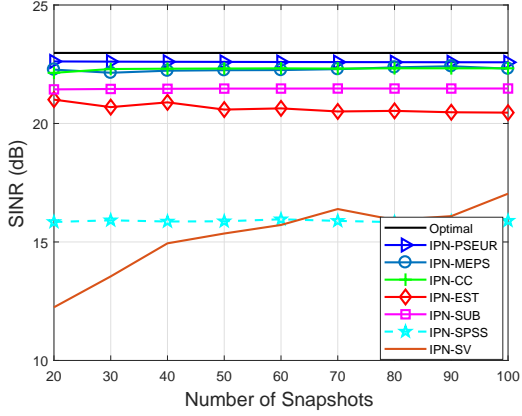}
\vspace{-0.75em}
    \caption{Output SINR versus Number of Snapshots for Example 1}
    \label{SINR_Mw_No_Mismatch}
\end{figure}

In this scenario, we examine the case when there is no mismatch between the assumed steering vector and the true one. The output SINR of the examined algorithms versus the SNR for a given training data size of 30 is shown in Fig.~\ref{SINR_SNR_No_Mismatch}. Since results for all SNRs are not distinguishable, the deviations from the optimal SINR are displayed in Fig.~\ref{SINR_Dev_No_Mismatch}. The performance of the proposed beamformer always achieves close to optimal values for a wide SNR range and outperforms the other beamformers. The output SINR of the IPN-SUB method is degraded more when the input SNR is $<-10$ dB. Also, the IPN-SPSS method has the lowest output SINR  for all range of SNRs, since it is sensitive to the number sensors. Fig.~\ref{SINR_Mw_No_Mismatch} depicts the output SINR of the tested methods for fixed SNR=10 dB while the number of snapshots is varied. Although the IPN-MEPS and IPN-CC beamformers show a good performance, the performance of the proposed beamformer is  better than the other tested beamformers.

\subsection{Example 2: Random signal look direction mismatch}
\begin{figure}[h]
    \centering
    \includegraphics[height=2.3in]{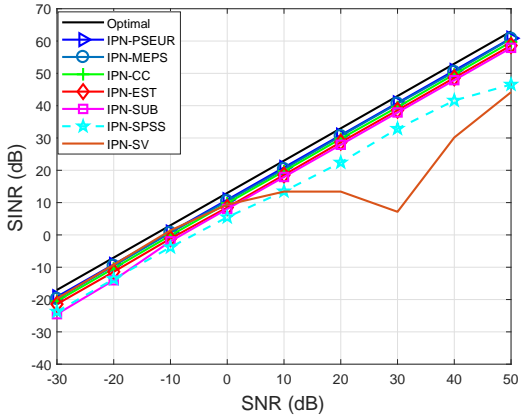}
\vspace{-0.75em}
    \caption{Output SINR versus SNR for Example 2}
    \label{SINR_SNR_Look_Mismatch}
\end{figure}

\begin{figure}[h]
    \centering
    \includegraphics[height=2.3in]{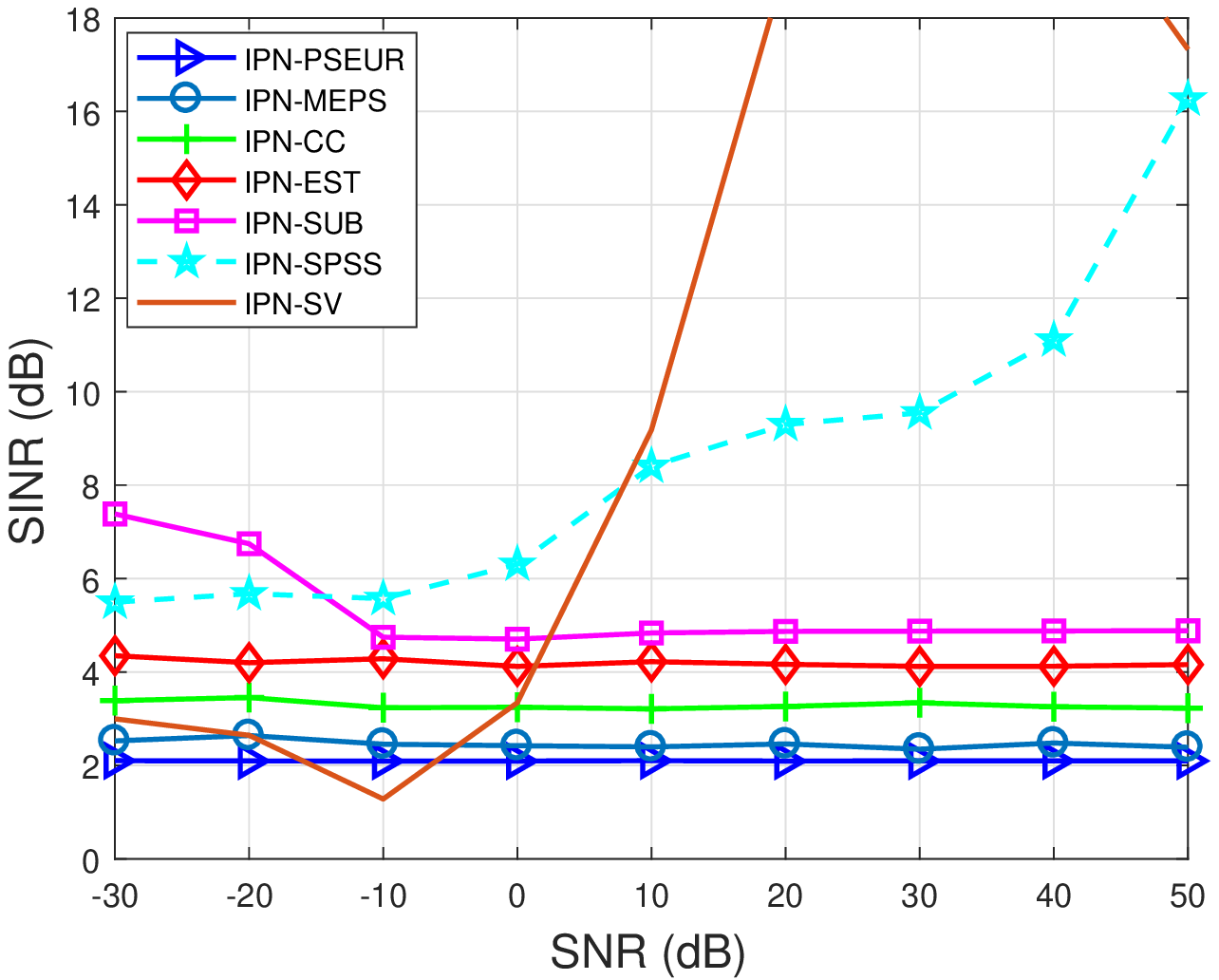}
\vspace{-0.75em}
    \caption{ Deviation from optimal SINR versus SNR for Example 2 }
    \label{SINR_Dev_Look_Mismatch}
\end{figure}

\begin{figure}[h]
    \centering
    \includegraphics[height=2.3in]{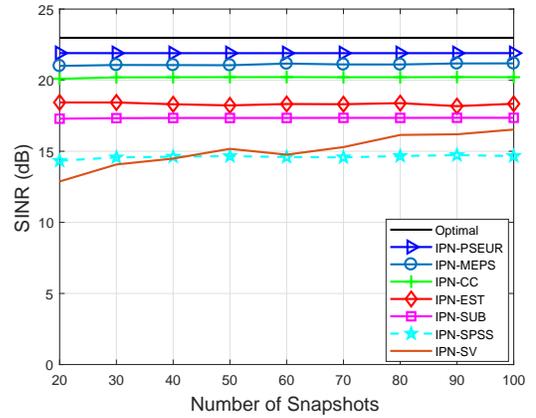}
\vspace{-0.75em}
    \caption{Output SINR versus Number of Snapshots for Example 2}
    \label{SINR_Mw_Look_Mismatch}
\end{figure}

In this scenario, we assume that the random direction mismatch of the desired signal and interferences are subject to uniform distribution in $[-5^o, 5^o]$ for each simulation run, which means that the direction of the SOI is varied between $[5^o,15^o]$ and the directions of two interferences are changed by
$[-55^o,-45^o]$ and $[25^o,35^o]$, respectively. Fig.~\ref{SINR_SNR_Look_Mismatch} shows the performance curves versus the input SNR while Fig.~\ref{SINR_Dev_Look_Mismatch} depicts the deviation between optimum and the tested beamformers. Fig.~\ref{SINR_SNR_Look_Mismatch} shows that the proposed beamformer has an advantage compared to the other tested beamformers in the case of signal look-direction
error at low and high SNRs. The performance of the IPN-SUB is degraded in the presence of look direction mismatch while the IPN-SV has good performance only up to 0 dB. Moreover, the IPN-SPSS method mantains its performance compared to the first scenario. The performance of the beamformers as the number of snapshots $N$ increases, is illustrated in Fig.~\ref{SINR_Mw_Look_Mismatch}.
In this scenario, the IPN-MEPS and IPN-CC methods demonstrate almost the same performance while the proposed approach exhibits better performance in the case of signal look direction errors.

\subsection{Example 3: Gain and phase perturbations error}
\begin{figure}[h]
    \centering
    \includegraphics[height=2.3in]{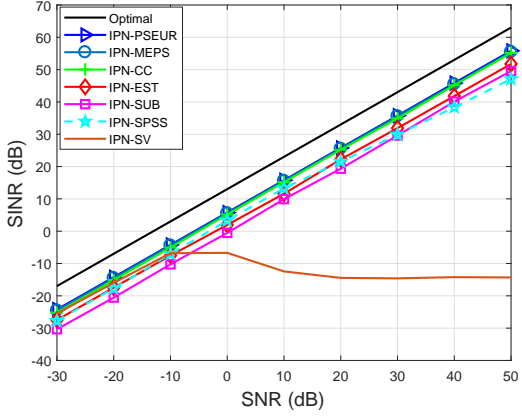}
\vspace{-0.75em}
    \caption{Output SINR versus SNR for Example 3}
    \label{SINR_SNR_Gain_Mismatch}
\end{figure}

\begin{figure}[h]
    \centering
    \includegraphics[height=2.3in]{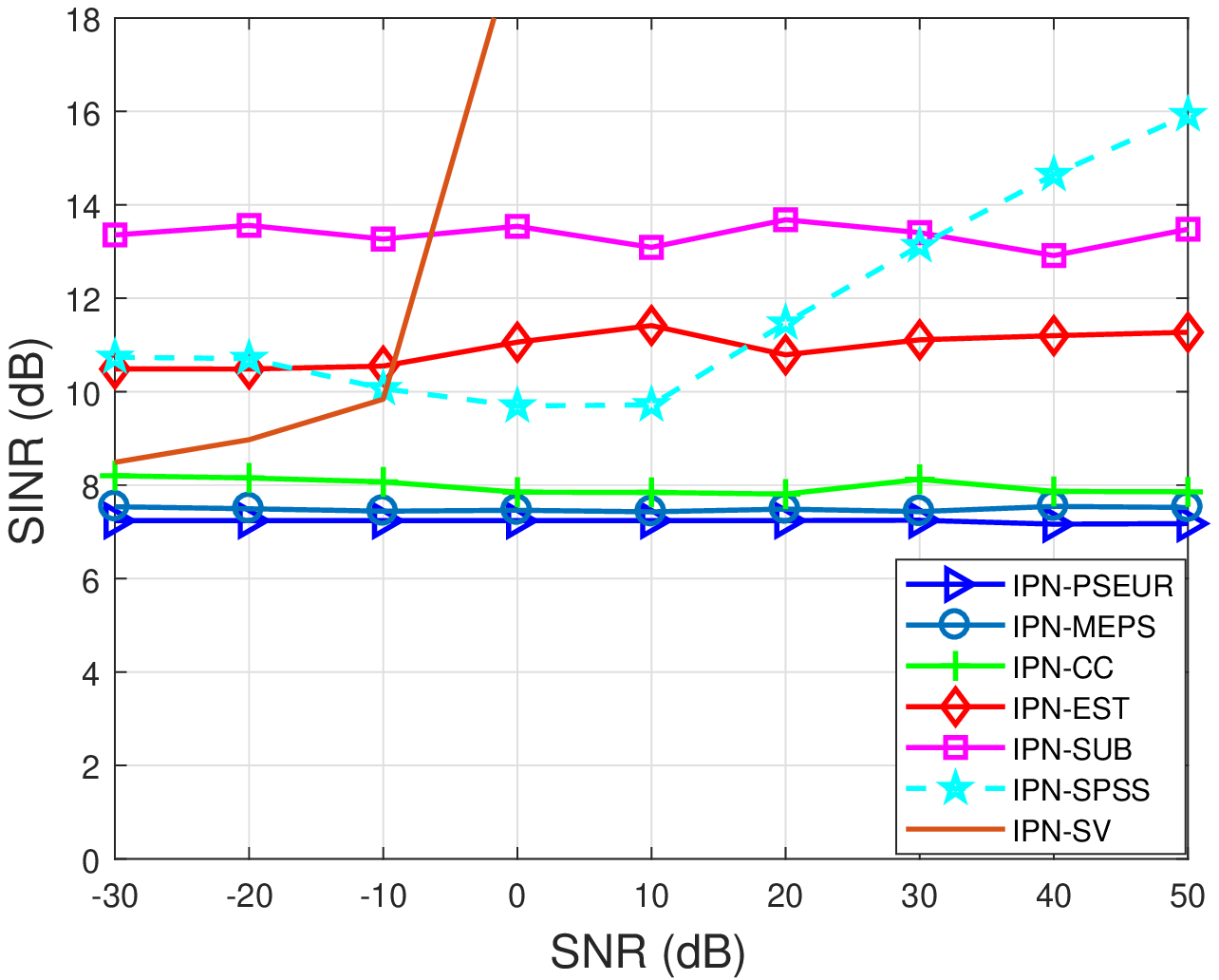}
\vspace{-0.75em}
    \caption{ Deviation from optimal SINR versus SNR for Example 3}
    \label{SINR_Dev_Gain_Mismatch}
\end{figure}

\begin{figure}[h]
    \centering
    \includegraphics[height=2.3in]{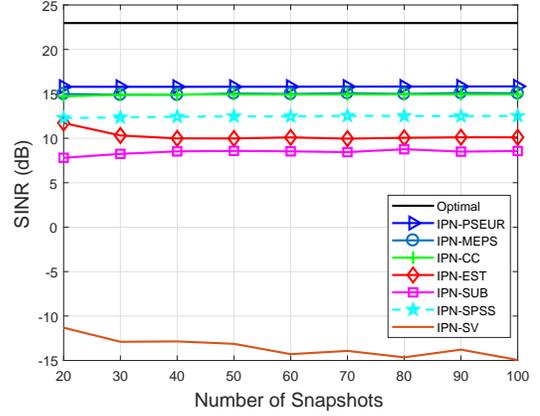}
\vspace{-0.75em}
    \caption{Output SINR versus Number of Snapshots for Example 3}
    \label{SINR_Mw_Gain_Mismatch}
\end{figure}

In this example, gain and phase mismatches on the array is described as
\begin{align}
    \mathbf{a}_m(\theta)=(1+\alpha_m) e^{j(\pi \sin \theta(m-1)+\beta_m)},
\end{align}
where $\alpha_m$ denotes  zero-mean random amplitude error at the $m$th sensor with a $\mathcal{N}(0, 0.05^2)$ distribution, and $\beta_m$ denotes the zero-mean phase error with distribution $\mathcal{N}(0, (0.025 \pi)^2)$. The output SINR of the tested beamformers versus the SNR for a fixed number of snapshots is represented in Fig.~\ref{SINR_SNR_Gain_Mismatch}. In order to illustrate clear results, we exhibit the deviation between the tested methods and the optimum SINR versus the input SNR in Fig.~\ref{SINR_Dev_Gain_Mismatch}.\\ From these results it can be seen that, although, the IPN-MEPS and IPN-CC beamformers have effective performance against  gain and phase errors, the proposed method has the most stable performance among the tested methods and almost reaches the optimal SINR, which assumes perfect knowledge of the
interference powers and the IPN covariance matrix. The IPN-SV beamformer performance is degraded compared to the random look direction mismatch example, since it is more sensitive to gain and phase perturbations, while the performance of the IPN-SPSS has improved. Meanwhile, Fig.~\ref{SINR_Mw_Gain_Mismatch} shows the stable performance for all the tested beamformers when the number of snapshots increased, and the proposed method has an output SINR that is closer to that of the optimal one.

\subsection{Example 4: Mismatch due to coherent local scattering}
\begin{figure}[!]
    \centering
    \includegraphics[height=2.3in]{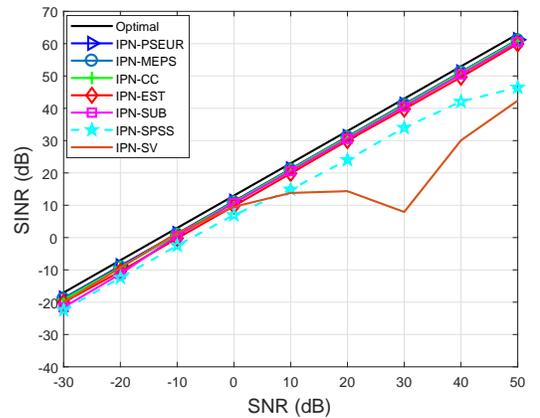}
\vspace{-0.75em}
    \caption{Output SINR versus SNR for Example 4}
    \label{SINR_SNR_Coherent}
\end{figure}

\begin{figure}[!]
    \centering
    \includegraphics[height=2.3in]{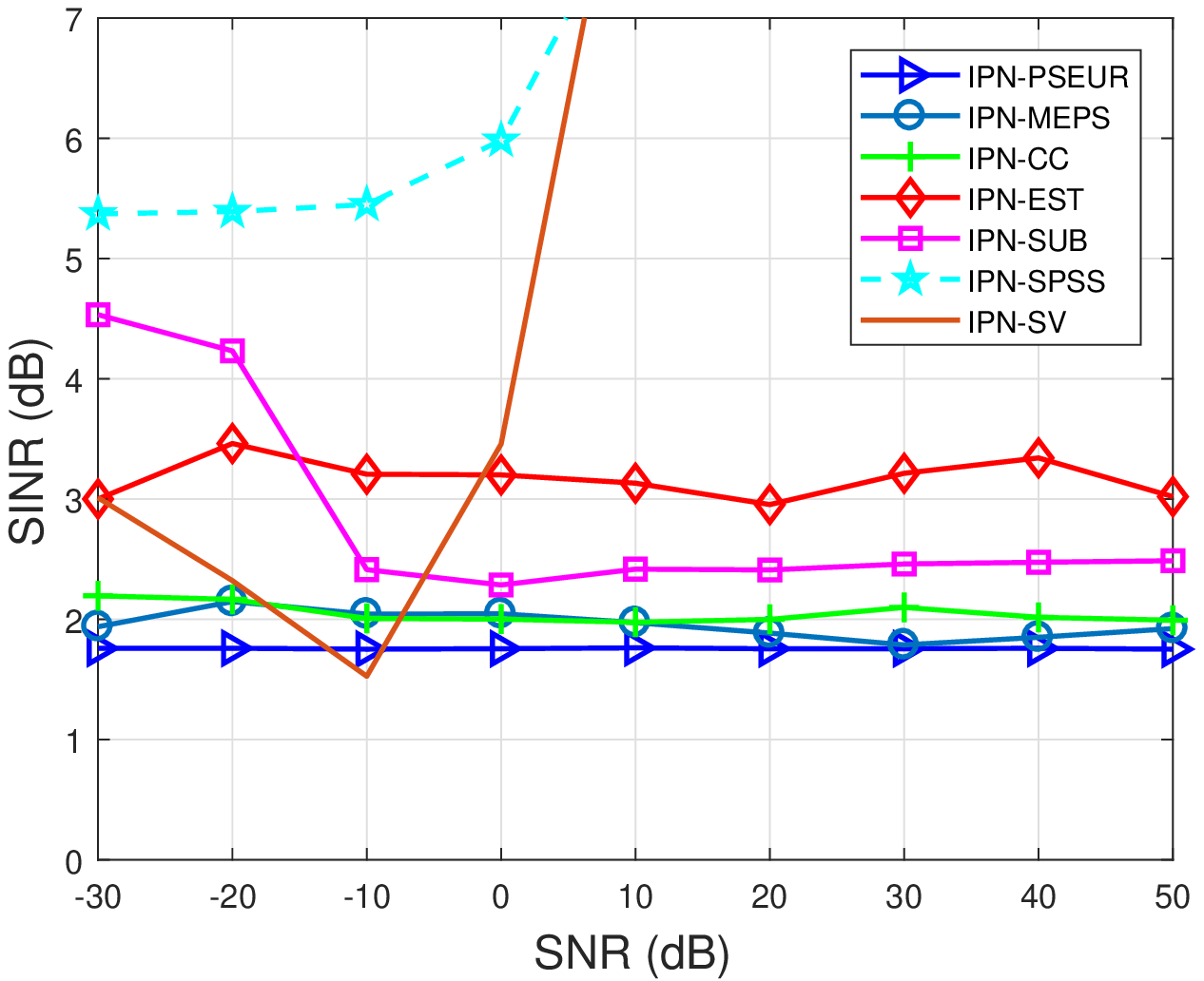}
\vspace{-0.75em}
    \caption{ Deviation from optimal SINR versus SNR for Example 4 }
    \label{SINR_Dev_Coherent}
\end{figure}

In the fourth scenario, the impact of the desired signal steering vector mismatch due to coherent local scattering
on array output SINR is considered. In this example, the presumed signal is a plane wave impinging from $ \hat{\theta}_1=10^\circ $, whereas the actual spatial signature is formed by five signal paths as
\begin{align}
    \tilde{\textbf{a}}=\mathbf{a}(\hat{\theta}_{1})+\sum_{i=1}^{4}e^{j\varphi_{i}} \textbf{d}(\theta_{i}),
\end{align}
where $ \mathbf{a}(\hat{\theta}_{1}) $ is the direct path and corresponds to the assumed signal steering vector, and $ \textbf{d}(\theta_{i}) $ represents the \textit{i}th coherently scattered path with the direction $ \theta_{i} $, (\textit{i}=1,2,3,4) which are randomly distributed in a Gaussian distribution with mean $\bar{\theta}_1 $ and standard deviation $ 2^\circ $. Also, the parameters $ \varphi_{i} $ denote the path phases which are drawn uniformly from the interval $ [0,2\pi]$ in each simulation run. Note that $ \theta_{i} $ and $ \varphi_{i} $ (\textit{i}=1,2,3,4) only change from run to run while remaining fixed from snapshot to snapshot. Assuming that the number of snapshots is fixed at 30, the output SINR versus the input SNR is demonstrated in Fig.~\ref{SINR_SNR_Coherent} while the deviation of all algorithms versus the optimal SINR is depicted in Fig.~\ref{SINR_Dev_Coherent}. It is evident that the IPN-SUB beamformer is sensitive to low SNRs while the IPN-CC and IPN-MEPS beamformers provide almost stable results. The proposed IPN-PSEUR algorithm is closest to the optimal output SINR among all algorithms.
\begin{figure}[!]
    \centering
    \includegraphics[height=2.3in]{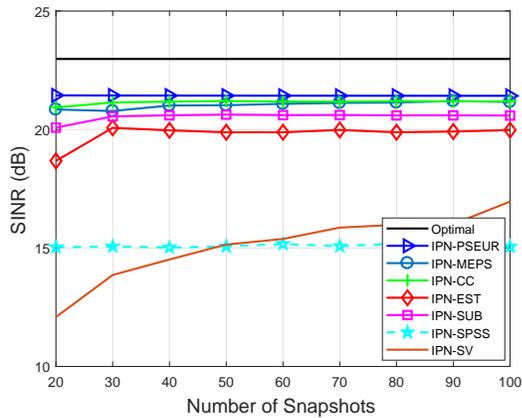}
\vspace{-0.75em}
    \caption{Output SINR versus Number of Snapshots for Example 4}
    \label{SINR_Mw_Coherent}
\end{figure}


\section{Conclusion} \label{Conclusion}
In this work, we proposed an effective adaptive beamforming approach based on the IPN covariance matrix that utilizes power spectral estimation and the uncertainty region around the IPN component. An efficient and adaptive method for estimating the uncertainty region where the null interference should be in every snapshot has been proposed. Furthermore, a new method to estimate the power spectrum of the interference and noise components was introduced, which is based on a piecewise constant function with only two levels to shape the directional response of the beamformer. Simulation results show that the proposed PSEUR algorithm outperforms recently reported approaches.

\ifCLASSOPTIONcaptionsoff
  \newpage
\fi

\bibliographystyle{IEEEtran}
\bibliography{Reference}

\begin{thebibliography}{10}
\providecommand{\url}[1]{#1}
\csname url@samestyle\endcsname
\providecommand{\newblock}{\relax}
\providecommand{\bibinfo}[2]{#2}
\providecommand{\BIBentrySTDinterwordspacing}{\spaceskip=0pt\relax}
\providecommand{\BIBentryALTinterwordstretchfactor}{4}
\providecommand{\BIBentryALTinterwordspacing}{\spaceskip=\fontdimen2\font plus
\BIBentryALTinterwordstretchfactor\fontdimen3\font minus
  \fontdimen4\font\relax}
\providecommand{\BIBforeignlanguage}[2]{{%
\expandafter\ifx\csname l@#1\endcsname\relax
\typeout{** WARNING: IEEEtran.bst: No hyphenation pattern has been}%
\typeout{** loaded for the language `#1'. Using the pattern for}%
\typeout{** the default language instead.}%
\else
\language=\csname l@#1\endcsname
\fi
#2}}
\providecommand{\BIBdecl}{\relax}
\BIBdecl

\bibitem{godara1997application}
L.~C. Godara, ``Application of antenna arrays to mobile communications. ii.
  beam-forming and direction-of-arrival considerations,'' \emph{Proceedings of
  the IEEE}, vol.~85, no.~8, pp. 1195--1245, 1997.

\bibitem{van2004detection}
H.~L. Van~Trees, \emph{Detection, Estimation, and Modulation Theory}.\hskip 1em
  plus 0.5em minus 0.4em\relax John Wiley \& Sons, New York, 2004.

\bibitem{capon1969high}
J.~Capon, ``High-resolution frequency-wavenumber spectrum analysis,''
  \emph{Proceedings of the IEEE}, vol.~57, no.~8, pp. 1408--1418, 1969.

\bibitem{li2003robust}
J.~Li, P.~Stoica, and Z.~Wang, ``On robust {Capon} beamforming and diagonal
  loading,'' \emph{IEEE Transactions on Signal Processing.}, vol.~51, no.~7,
  pp. 1702--1715, 2003.

\bibitem{elnashar2006further}
A.~Elnashar, S.~M. Elnoubi, and H.~A. El-Mikati, ``Further study on robust
  adaptive beamforming with optimum diagonal loading,'' \emph{IEEE Transactions
  on Antennas and Propagation}, vol.~54, no.~12, pp. 3647--3658, 2006.

\bibitem{kukrer2014generalised}
O.~Kukrer and S.~Mohammadzadeh, ``Generalised loading algorithm for adaptive
  beamforming in ulas,'' \emph{Electronics Letters}, vol.~50, no.~13, pp.
  910--912, 2014.

\bibitem{rstap}
X.~Wang, Z.~Yang, J.~Huang, and R.~C. de~Lamare, ``Robust two-stage
  reduced-dimension sparsity-aware stap for airborne radar with coprime
  arrays,'' \emph{IEEE Transactions on Signal Processing}, vol.~68, pp. 81--96,
  2020.

\bibitem{rcfprec}
V.~M.~T. Palhares, A.~R. Flores, and R.~C. de~Lamare, ``Robust mmse precoding
  and power allocation for cell-free massive mimo systems,'' \emph{IEEE
  Transactions on Vehicular Technology}, vol.~70, no.~5, pp. 5115--5120, 2021.

\bibitem{vorobyov2003robust}
S.~A. Vorobyov, A.~B. Gershman, and Z.-Q. Luo, ``Robust adaptive beamforming
  using worst-case performance optimization: A solution to the signal mismatch
  problem,'' \emph{IEEE Transactions on Signal Processing}, vol.~51, no.~2, pp.
  313--324, 2003.

\bibitem{yu2010robust}
Z.~L. Yu, Z.~Gu, J.~Zhou, Y.~Li, W.~Ser, and M.~H. Er, ``A robust adaptive
  beamformer based on worst-case semi-definite programming,'' \emph{IEEE
  Transactions on Signal Processing}, vol.~58, no.~11, pp. 5914--5919, 2010.

\bibitem{huang2012modified}
F.~Huang, W.~Sheng, and X.~Ma, ``Modified projection approach for robust
  adaptive array beamforming,'' \emph{Signal Processing}, vol.~92, no.~7, pp.
  1758--1763, 2012.

\bibitem{gbd}
K.~Zu, R.~C. de~Lamare, and M.~Haardt, ``Generalized design of low-complexity
  block diagonalization type precoding algorithms for multiuser mimo systems,''
  \emph{IEEE Transactions on Communications}, vol.~61, no.~10, pp. 4232--4242,
  2013.

\bibitem{wlbd}
W.~Zhang, R.~C. de~Lamare, C.~Pan, M.~Chen, J.~Dai, B.~Wu, and X.~Bao, ``Widely
  linear precoding for large-scale mimo with iqi: Algorithms and performance
  analysis,'' \emph{IEEE Transactions on Wireless Communications}, vol.~16,
  no.~5, pp. 3298--3312, 2017.

\bibitem{rsbd}
A.~R. Flores, R.~C. de~Lamare, and B.~Clerckx, ``Linear precoding and stream
  combining for rate splitting in multiuser mimo systems,'' \emph{IEEE
  Communications Letters}, vol.~24, no.~4, pp. 890--894, 2020.

\bibitem{mohammadzadeh2018modified}
S.~Mohammadzadeh and O.~Kukrer, ``Modified robust {Capon} beamforming with
  approximate orthogonal projection onto the signal-plus-interference
  subspace,'' \emph{Circuits, Systems, and Signal Processing}, pp. 1--18, 2018.

\bibitem{mohammadzadeh2022letter}
S.~Mohammadzadeh, V.~H. Nascimento, R.~C. de~Lamare, and N.~Hajarolasvadi,
  ``Robust beamforming based on complex-valued convolutional neural networks
  for sensor arrays,'' \emph{IEEE Signal Processing Letters}, 2022.

\bibitem{gu2012robust}
Y.~Gu and A.~Leshem, ``Robust adaptive beamforming based on interference
  covariance matrix reconstruction and steering vector estimation,'' \emph{IEEE
  Transactions on Signal Processing}, vol.~60, no.~7, pp. 3881--3885, 2012.

\bibitem{chen2015robust}
F.~Chen, F.~Shen, and J.~Song, ``Robust adaptive beamforming using
  low-complexity correlation coefficient calculation algorithms,''
  \emph{Electronics Letters}, vol.~51, no.~6, pp. 443--445, 2015.

\bibitem{gu2014robust}
Y.~Gu, N.~A. Goodman, S.~Hong, and Y.~Li, ``Robust adaptive beamforming based
  on interference covariance matrix sparse reconstruction,'' \emph{Signal
  Processing}, vol.~96, pp. 375--381, 2014.

\bibitem{ruan2014robust}
H.~Ruan and R.~C. de~Lamare, ``Robust adaptive beamforming using a
  low-complexity shrinkage-based mismatch estimation algorithm.'' \emph{IEEE
  Signal Processing Letters.}, vol.~21, no.~1, pp. 60--64, 2014.

\bibitem{ruan2016}
H.~{Ruan} and R.~C. {de Lamare}, ``Robust adaptive beamforming based on
  low-rank and cross-correlation techniques,'' \emph{IEEE Transactions on
  Signal Processing}, vol.~64, no.~15, pp. 3919--3932, Aug 2016.

\bibitem{lrcc}
H.~Ruan and R.~C. de~Lamare, ``Distributed robust beamforming based on low-rank
  and cross-correlation techniques: Design and analysis,'' \emph{IEEE
  Transactions on Signal Processing}, vol.~67, no.~24, pp. 6411--6423, 2019.

\bibitem{zhang2016interference}
Z.~Zhang, W.~Liu, W.~Leng, A.~Wang, and H.~Shi, ``Interference-plus-noise
  covariance matrix reconstruction via spatial power spectrum sampling for
  robust adaptive beamforming,'' \emph{IEEE Signal Processing Letters},
  vol.~23, no.~1, pp. 121--125, 2016.

\bibitem{yuan2017robust}
X.~Yuan and L.~Gan, ``Robust adaptive beamforming via a novel subspace method
  for interference covariance matrix reconstruction,'' \emph{Signal
  Processing}, vol. 130, pp. 233--242, 2017.

\bibitem{zheng2018covariance}
Z.~Zheng, Y.~Zheng, W.-Q. Wang, and H.~Zhang, ``Covariance matrix
  reconstruction with interference steering vector and power estimation for
  robust adaptive beamforming,'' \emph{IEEE Transactions on Vehicular
  Technology}, vol.~67, no.~9, pp. 8495--8503, 2018.

\bibitem{mohammadzadeh2018adaptive}
S.~Mohammadzadeh and O.~Kukrer, ``Adaptive beamforming based on theoretical
  interference-plus-noise covariance and direction-of-arrival estimation,''
  \emph{IET Signal Processing}, vol.~12, no.~7, pp. 819--825, 2018.

\bibitem{zheng2019robustt}
Z.~Zheng, T.~Yang, W.-Q. Wang, and H.~C. So, ``Robust adaptive beamforming via
  simplified interference power estimation,'' \emph{IEEE Transactions on
  Aerospace and Electronic Systems}, vol.~55, no.~6, pp. 3139--3152, 2019.

\bibitem{zheng2019robust}
Z.~Zheng, W.-Q. Wang, H.~C. So, and Y.~Liao, ``Robust adaptive beamforming
  using a novel signal power estimation algorithm,'' \emph{Digital Signal
  Processing}, vol.~95, p. 102574, 2019.

\bibitem{chen2018adaptive}
P.~Chen, Y.~Yang, Y.~Wang, and Y.~Ma, ``Adaptive beamforming with sensor
  position errors using covariance matrix construction based on subspace bases
  transition,'' \emph{IEEE Signal Processing Letters}, vol.~26, no.~1, pp.
  19--23, 2019.

\bibitem{MEPSalgorithm}
S.~Mohammadzadeh, V.~H. Nascimento, R.~C. de~Lamare, and O.~Kukrer, ``Maximum
  entropy-based interference-plus-noise covariance matrix reconstruction for
  robust adaptive beamforming,'' \emph{IEEE Signal Processing Letters.},
  vol.~27, pp. 845--849, 2020.

\bibitem{mohammadzadeh2020low}
------, ``Low-cost adaptive maximum entropy covariance matrix reconstruction
  for robust beamforming,'' in \emph{2020 54th Asilomar Conference on Signals,
  Systems, and Computers}.\hskip 1em plus 0.5em minus 0.4em\relax IEEE, 2020,
  pp. 1462--1466.

\bibitem{zhu2020robust}
X.~Zhu, X.~Xu, and Z.~Ye, ``Robust adaptive beamforming via subspace for
  interference covariance matrix reconstruction,'' \emph{Signal Processing},
  vol. 167, p. 107289, 2020.

\bibitem{yang2021robust}
H.~Yang, P.~Wang, and Z.~Ye, ``Robust adaptive beamforming via covariance
  matrix reconstruction under colored noise,'' \emph{IEEE Signal Processing
  Letters}, vol.~28, pp. 1759--1763, 2021.

\bibitem{sun2021robust}
S.~Sun and Z.~Ye, ``Robust adaptive beamforming based on a method for steering
  vector estimation and interference covariance matrix reconstruction,''
  \emph{Signal Processing}, vol. 182, p. 107939, 2021.

\bibitem{mohammadzadeh2021robust}
S.~Mohammadzadeh, V.~H. Nascimento, R.~C. de~Lamare, and O.~Kukrer, ``Robust
  adaptive beamforming based on virtual sensors using low-complexity spatial
  sampling,'' \emph{Signal Processing}, vol. 188, p. 108172, 2021.

\bibitem{mohammadzadeh2022robust}
S.~Mohammadzadeh, V.~H. Nascimento, R.~C. De~Lamare, and O.~Kukrer, ``Robust
  adaptive beamforming based on power method processing and spatial spectrum
  matching,'' in \emph{ICASSP 2022-2022 IEEE International Conference on
  Acoustics, Speech and Signal Processing (ICASSP)}.\hskip 1em plus 0.5em minus
  0.4em\relax IEEE, 2022, pp. 4903--4907.

\bibitem{spa}
R.~C. De~Lamare and R.~Sampaio-Neto, ``Minimum mean-squared error iterative
  successive parallel arbitrated decision feedback detectors for ds-cdma
  systems,'' \emph{IEEE Transactions on Communications}, vol.~56, no.~5, pp.
  778--789, 2008.

\bibitem{jio}
R.~C. de~Lamare and R.~Sampaio-Neto, ``Reduced-rank adaptive filtering based on
  joint iterative optimization of adaptive filters,'' \emph{IEEE Signal
  Processing Letters}, vol.~14, no.~12, pp. 980--983, 2007.

\bibitem{jidf}
------, ``Adaptive reduced-rank processing based on joint and iterative
  interpolation, decimation, and filtering,'' \emph{IEEE Transactions on Signal
  Processing}, vol.~57, no.~7, pp. 2503--2514, 2009.

\bibitem{jiodoa}
L.~Wang, R.~C. de~Lamare, and M.~Haardt, ``Direction finding algorithms based
  on joint iterative subspace optimization,'' \emph{IEEE Transactions on
  Aerospace and Electronic Systems}, vol.~50, no.~4, pp. 2541--2553, 2014.

\bibitem{alrdoa}
L.~Qiu, Y.~Cai, R.~C. de~Lamare, and M.~Zhao, ``Reduced-rank doa estimation
  algorithms based on alternating low-rank decomposition,'' \emph{IEEE Signal
  Processing Letters}, vol.~23, no.~5, pp. 565--569, 2016.

\bibitem{mskaesprit}
S.~F.~B. Pinto and R.~C. de~Lamare, ``Multistep knowledge-aided iterative
  esprit: Design and analysis,'' \emph{IEEE Transactions on Aerospace and
  Electronic Systems}, vol.~54, no.~5, pp. 2189--2201, 2018.

\bibitem{christensen2009sinusoidal}
M.~G. Christensen, A.~Jakobsson, and S.~H. Jensen, ``Sinusoidal order
  estimation using angles between subspaces,'' \emph{EURASIP Journal on
  Advances in Signal Processing}, vol. 2009, pp. 1--11, 2009.

\bibitem{mohammadzadeh2019robust}
S.~Mohammadzadeh and O.~Kukrer, ``Robust adaptive beamforming for fast moving
  interference based on the covariance matrix reconstruction,'' \emph{IET
  Signal Processing}, vol.~13, no.~4, pp. 486--493, 2019.

\bibitem{lie2011adaptive}
J.~P. Lie, W.~Ser, and C.~M.~S. See, ``Adaptive uncertainty based iterative
  robust {Capon} beamformer using steering vector mismatch estimation,''
  \emph{IEEE Transactions on Signal Processing}, vol.~59, no.~9, pp.
  4483--4488, 2011.

\bibitem{zhang2013robust}
W.~Zhang, J.~Wang, and S.~Wu, ``Robust capon beamforming against large doa
  mismatch,'' \emph{Signal Processing}, vol.~93, no.~4, pp. 804--810, 2013.

\bibitem{khabbazibasmenj2012robust}
A.~Khabbazibasmenj, S.~A. Vorobyov, and A.~Hassanien, ``Robust adaptive
  beamforming based on steering vector estimation with as little as possible
  prior information,'' \emph{IEEE Transactions on Signal Processing}, vol.~60,
  no.~6, pp. 2974--2987, 2012.

\bibitem{grant2008cvx}
M.~Grant, S.~Boyd, and Y.~Ye, ``Cvx: Matlab software for disciplined convex
  programming,'' 2008.

\end{thebibliography}

\end{document}